\documentclass[aip, rsi,amsmath,amssymb,reprint]{revtex4-2}

\usepackage{graphicx}
\usepackage{dcolumn}
\usepackage{bm}
\usepackage[utf8]{inputenc}
\usepackage[T1]{fontenc}
\usepackage{mathptmx}
\usepackage{etoolbox}
\usepackage[binary-units]{siunitx}[=v2]
\DeclareSIUnit{\count}{counts}
\usepackage{braket}

\makeatletter
\def\@email#1#2{%
	\endgroup
	\patchcmd{\titleblock@produce}
	{\frontmatter@RRAPformat}
	{\frontmatter@RRAPformat{\produce@RRAP{*#1\href{mailto:#2}{#2}}}\frontmatter@RRAPformat}
	{}{}
}%
\makeatother
\begin{document}
	
	\title[Fast Laser Switching and Chirping]{Switching, Amplifying, and Chirping Diode Lasers with Current Pulses for High Bandwidth Quantum Technologies}
	\author{Gianni Buser}
	\email{gianni.buser@unibas.ch}
	
	\affiliation{University of Basel, Department of Physics, Klingelbergstrasse 82, 4056 Basel, Switzerland}%
	
	\date{\today}
	
	\begin{abstract}
		A series of simple and low-cost devices for switching, amplifying, and chirping diode lasers based on current modulation are presented. Direct modulation of diode laser currents is rarely sufficient to establish precise amplitude and phase control over light, as its effects on these parameters are not independent. These devices overcome this limitation by exploiting amplifier saturation and dramatically outperform commonly used external modulators in key figures of merit for quantum technological applications. Semiconductor optical amplifiers operated on either rubidium D line are recast as intensity switches and shown to achieve ON:OFF ratios $>\num{e6}$ in as little as \SI{50}{\nano\second}. Current is switched to a \SI{795}{\nano\meter} wavelength (Rb D$_1$) tapered amplifier to produce optical pulses of few nanosecond duration and peak powers of \SI{3}{\watt} at a similar extinction ratio. Fast rf pulses are applied directly to a laser diode to shift its emission frequency by up to \SI{300}{\mega\hertz} in either direction and at a maximum chirp rate of \SI{150}{\mega\hertz\per\nano\second}. Finally, the latter components are combined, yielding a system that produces watt-level optical pulses with arbitrary frequency chirps in the given range and $<\SI{2}{\percent}$ residual intensity variation, all within \SI{65}{\nano\second} upon asynchronous demand. Such systems have broad application in atomic, molecular, and optical physics, and are of particular interest to fast experiments simultaneously requiring high power and low noise, for example quantum memory experiments with atomic vapors.  
	\end{abstract}
	
	\maketitle
	
	\section{Introduction}
		
	Rapid control of the amplitude and frequency of laser light fields is critical to many of their key applications, including optical communication \cite{Sibley2020}, optical computing \cite{McMahon2023}, and control over atomic or molecular matter \cite{Torosov2021}. In the domain of quantum optical technologies, the signals of interest are single photons or weak coherent pulses with intensities on the single photon level \cite{Scarani2009, Xu2020, Slussarenko2019, Pirandola2018}. Here the nanosecond-timing, gigahertz-bandwidth regime is intrinsically attractive to applications because high rates are technologically useful. Simultaneously, it is home to high quality single-emitter photon sources \cite{Aharonovich2016,Senellart2017}, including ones suited for hybrid atomic interfaces such as Rb-like quantum dots \cite{Zhai2022}. 
	
	A serious challenge in fast quantum optics experiments is the discrimination of these weak photonic signals from intense laser fields. Diverse and useful non-linear optical processes, including quantum frequency conversion by sum- or difference frequency generation \cite{Ma2012} in the context of transduction\cite{Lauk2020}, (entangled) photon pair production by parametric down conversion\cite{Kwiat1995}, and an abundance of schemes facilitating quantum light-matter interactions\cite{Hammerer2010} all place high power ``pumping'' or ``control'' beams into spatial modes shared by single photon signals. This puts high demands both on the speed and dynamic range of laser control, and when the latter is insufficient, this must be compensated by filtering leading to photon loss. Simply put, commonly used optical switches for external laser modulation, including ubiquitous acousto- and electro-optic modulators (AOMs and EOMs), do not reach intensity extinction ratios (IER) concordant with the notion of a laser being OFF in nanoseconds when the signal of interest is on the single photon level. Comparing energies, the emissions of a \si{\milli\watt}--\si{\watt} laser in \SI{1}{\nano\second} contain $E=\si{\pico\joule}$--$\si{\nano\joule}$, whereas a photon at \SI{795}{\nano\meter} contains only $E_{\text{ph}}\approx\SI{2.5e-19}{\joule}$. Desirable IER may thus exceed even \SI{100}{\deci\bel}. 

	\begin{table}
	\begin{ruledtabular}
		\begin{tabular}{ccccc}
			Switch&$\tau_d$ [s]&$\tau_s$ [s]&IER&$P_\text{max}$ [W]\\ 
			\hline
			\vspace{-9pt}\\
			Shutter & \num{e-4} & \num{e-4} & $\infty$ & >\num{e2}\\
			AOM&\num{e-6}&\num{e-5}&>\SI{80}{\deci\bel}&\num{e1}\\
			Waveguide EOM&\num{e-9}&\num{e-10}&\SI{30}{\deci\bel}&\num{e-3} \\
			Bulk EOM&\num{e-7}&\num{e-9}&\SI{30}{\deci\bel}&\num{e0}\\
			SOA&\num{e-8}&\num{e-8}&>\SI{60}{\deci\bel}&\num{e-2}\\
			TA&\num{e-8}&\num{e-8}&\SI{60}{\deci\bel}&\num{e0}
		\end{tabular}
	\end{ruledtabular}
	\caption{Comparison of speed and intensity performance of commonly used optical switches, by typical order of magnitude only. $\tau_d$ latency, i.e. the time between triggering a device and the beginning of an effect on the optical output, $\tau_s$ fall time, after the latency, to reach the specified extinction ratio, IER ratio of optical powers in ON and OFF states, $P_\text{max}$ maximum feasible continuous optical output in vis/NIR}
	\end{table}
	
	Mechanical shutters physically blocking a beam generally extinguish it completely. If the aperture is closed with a mirror redirecting the laser to an actively cooled beam trap, even kilowatt powers can be handled. Unfortunately, accelerating an object into a beam path is far slower than electronic means of switching. Almost as good in steady-state IER are AOMs, given sufficient electronic isolation and care in spatially separating the diffracted order from the direct transmission. Although these devices are often specified with nanosecond scale switching times, however, the measure at hand for such claims is a $90:10$ response. AOMs cannot achieve their steady-state IER until 10s of microseconds after being triggered. This delay is an inherent property of their principle of operation \cite{Sibley2020} -- there are limits on how fast sound waves can travel and dissipate -- not merely a technical limitation. In the visible and NIR, EOMs can switch intensities over about 3 orders of magnitude. Waveguide-based devices outperform their bulk counterparts in speed, but they tend to accumulate photorefractive damage \cite{Zhu2021} when operated even just at milliwatt powers. Higher IER can be achieved by cascading devices, but this compounds insertion loss reducing the available optical output. For a generic overview of typical performance across common devices see Table~1. Generally, modern, performant systems externally modulating both a laser's amplitude and phase sport considerable device overhead, and achieve what they set out to do in synchronous operation \cite{Clarke2022}. Typically, component latency is such that control within short times in response to asynchronous demand is simply not possible.
	
	Direct current modulation of laser diodes and optical amplifiers is a well-known alternative approach. The method is simple, cheap, and fast, routinely enabling tens of \si{\giga\bit\per\second} data rates when used to encode information in a telecommunications context \cite{Zhu2018}. Indeed combinations of bandwidth enhancing techniques have pushed direct modulation bandwidths beyond \SI{65}{\giga\hertz} in lasers without strong mode confinement \cite{Matsui2021}, and an even higher bandwidth of \SI{108}{\giga\hertz} has been shown in a membrane laser \cite{Yamaoka2020}. In quantum optical control applications, however, direct modulation is quite rare due to perceived downsides. The most prominent troubles are twofold. Primarily, modulation of a laser diode's current has a linked and modulation frequency dependent effect on both the amplitude and phase of the light\cite{Petermann1988}, which is at odds with independently varying these parameters. Secondarily, the response of the system must be characterized optically, as variations in the semiconductors preclude precise prior predictions of the output\cite{Amtmann2023}. Nevertheless, in this paper I make a case for the direct modulation approach, demonstrating detection limited IER beyond $\SI{60}{\deci\bel}$ achieved in 10s of nanoseconds by current switching continuously seeded semiconductor optical amplifiers (SOA) and tapered amplifiers (TA), as well as controlled pulse chirps at rates up to \SI{150}{\mega\hertz\per\nano\second} by applying rf pulses directly to laser diodes. For the latter, I minimize intensity fluctuations by passing the frequency modulated light through an optical amplifier in saturation, imprinting the frequency shifts on stable \SI{3}{\watt} peak-power pulsed output. Further, I find that given an initial characterization, diode response follows simple phenomenology enabling straightforward precision control.

	\section{Optical Amplifiers as Switches}
	
	Semiconductor optical amplifiers are a staple of optical switching in telecommunication network architectures \cite{Assadihaghi2010}. In this field, they are famous for exceptional dynamic range and high extinction ratios \cite{Tanaka2009}. This attribute lies close at hand. When current is passed through the device input seed light is amplified, with typical small signal gains on the order of \SI{30}{\deci\bel}. Simultaneously, in the absence of current, the semiconductor acts as a saturable absorber, yielding significant insertion loss rather than gain. In recent years devices with gain regions in the NIR compatible with alkali transitions have become commercially available, and the favorable properties that lead to their use in telecom translate seamlessly into the field of atom optics. Measured in the steady-state, a commercial, fiber-pigtailed SOA (Superlum SOA-332-DBUT-PM) configured to output \SI{20}{\milli\watt} of light on a rubidium D line (D1 at \SI{795}{\nano\meter}, D2 at \SI{780}{\nano\meter}) and utilizing its full gain range transmits no more than \SI{500}{\pico\watt} (measurement limit) of the incident seed light when no current is passed through it, implying a steady state IER $\geq\SI{75}{\deci\bel}$.
	
	Experiments utilizing laser light even for simple quantum control applications like state preparation can quickly become limited by the quality of optical switching in the high bandwidth regime \cite{Wolters2017, Thomas2019, Esguerra2023}. Particularly relevant in characterizing a switch for such uses is the IER as a function of time, as this is the critical metric for their use in realistic asynchronous applications. This characterization is demanding as it simultaneously requires a high time resolution to resolve the dynamics (beyond the typical performance of spectrum analyzers) and excellent dynamic range (beyond the typical performance of photo diodes and oscilloscopes) to cover many orders of magnitude in intensity. 

	\begin{figure}
	\includegraphics[width=\columnwidth]{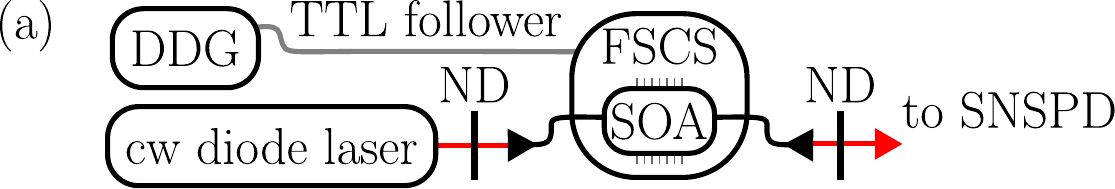}\\\vspace{0.2cm}
	\includegraphics[width=\columnwidth]{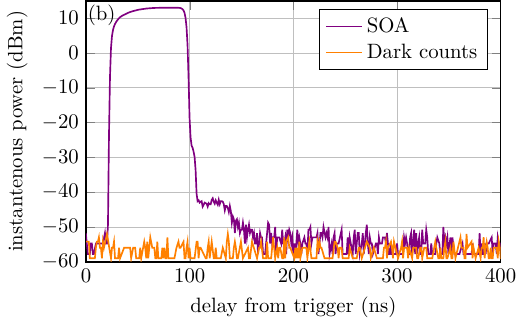}
	\caption{(a) Setup to characterize SOA switching performance. DDG digital delay generator, TTL transistor-transistor logic, FSCS (home built) fast switching current source, SOA semiconductor optical amplifier, ND neutral density, SNSPD superconducting nanowire single photon detector. (b) Time resolved optical switching behavior of a SOA under \SI{1}{\mega\hertz} pulsing of a forward current of \SI{168}{\milli\ampere} for \SI{75}{\nano\second}. The data are collected in $12\times\SI{81}{\pico\second}$ time bins, and the time axis is corrected for propagation and detection delay to correspond to the intensity at the SOA's pig-tailed fiber output. The device's maximal output power at \SI{795}{\nano\meter} is \SI{20}{\milli\watt}, which is used to calibrate the y-axis. Simultaneously, this light can be turned off over seven orders of magnitude within \SI{200}{\nano\second}, limited by the dark count set dynamic range of the detector.}
	\end{figure}
	
	A sketch of the characterization setup to determine the SOAs performance as an optical switch is shown in Fig.~1(a). To satisfy both timing and dynamic range demands, optical signals are measured using superconducting nanowire single photon detectors (SNSPD). An operating current of \SI{168}{\milli\ampere} is switched periodically to the SOA for \SI{75}{\nano\second} at a rate of \SI{1}{\mega\hertz}. Here and everywhere, logical signals are generated internally by a digital delay generator (Highland Technology T564). Response to an external trigger adds only \SI{20}{\nano\second} insertion delay and \SI{30}{\pico\second} jitter, which immediately readies all setups for asynchronous operation. The SOA is seeded with \SI{60(6)}{\micro\watt} of cw laser light, chosen to obtain the rated maximum of \SI{20}{\milli\watt} output power ($g=\SI{25.2(4)}{\deci\bel}$). This power is sufficient, for instance, to optically pump an ensemble of alkali atoms at saturation in a volume of radius \SI{1}{\centi\meter}. The light is attenuated to a level where, on average, one million photons are detected each second by the SNSPD (Single Quantum EOS, \SI{85}{\percent} system detection efficiency, \SI{20}{\pico\second} FWHM jitter, dark count rate <\SI{10}{\per\second}). Detection events are recorded with a time-tagger (quTools quTau, \SI{81}{\pico\second} time resolution). The switching behavior of this current controlled SOA is shown in Fig.~1(b), in a measurement integrated for \SI{12.7}{\minute} to resolve the y-axis as well as possible. The known maximum optical output is used to calibrate the y-axis into units of power. The output falls on a steep flank, reaching a suppression level of \SI{-56}{\deci\bel} in \SI{17}{\nano\second} from the start of switching, and over \SI{-60}{\deci\bel} within \SI{50}{\nano\second}. A direct comparison reveals that after this point the measured suppression is limited by dark counts, recorded separately without any light incident over the same integration time for comparison. The observed average dark count rate is only \SI{1.2}{\count\per\second}, considerably better than bare instrument specification. This is achieved by turning down the bias current to the nanowire well below its critical point. Dark count rates in SNSPDs can be illumination dependent due to thermal radiation from the fibers, or in more complicated ways too\cite{Chen2015}, and some care must be taken to ensure that only their intrinsic rate limits a measurement.
	
	In a separate characterization with careful compensation of cables, the latency (i.e. the time between an electronic trigger input and the start of an optical response) of the device is measured to be \SI{23}{\nano\second}, and full output is reached within \SI{45}{\nano\second} from the beginning of its response. The electronic driver powering the switch is home built and its current output follows a transistor-transistor logic (TTL) signal with no restrictions on cycle time, but it does limit the switching flank steepness. The driver redirects current between the SOA and a dispersive dummy load mimicking the SOA using a MOSFET, with an electrical rise (fall) time of \SI{10}{\nano\second} (\SI{3}{\nano\second}). This limitation is verified experimentally using a faster pulsed current source (Highland Technology T160) suitable for low duty cycle operation. The ultimate limit, which I do not resolve, is expected to be a function of the photon lifetime in the semiconductor\cite{Sibley2020}. In the past couple of years, turnkey SOA/driver devices for switching applications at alkali wavelengths have been commercialized, simplifying their use even further. This choice of home built driver is made to suit the application of optical pumping, which can require large duty cycles. The switching performance is replicated over four tested devices with light on both rubidium D lines. Operation near the current threshold of the SOA ($g=1$) limits the maximum IER to about \SI{-45}{\deci\bel}, but this level is reached equally quickly. It is thus plausible, albeit just barely beyond the measurement limit of available devices, that the full steady state IER is reached within only \SI{60}{\nano\second} when the SOAs are operated at full gain. The output power obtained is a strict limitation, as higher powers induce catastrophic optical damage on the chip end facet where it is fiber pig-tailed. These devices have been used to switch optical pumping for years in our labs, and have enabled the asynchronous operation of diverse memory experiments \cite{Buser2022,Mottola2023}.
	
	To overcome the limitation on output power, the same switching principle is applied to a tapered amplifier (Eagleyard EYP-TPA-0795-02000-4006-BTU02-0000) rated to output \SI{2}{\watt} of \SI{795}{\nano\meter} light into free space. The current to this device, maximally \SI{4}{\ampere}, is switched by a commercial driver (Aerodiode CCS-HPP, \SI{4}{\ampere} maximum current, \SI{10}{\pico\second} (\SI{2}{\nano\second}) resolution for pulse output up to (longer than) \SI{10}{\nano\second}). It is used to generate intense laser pulses of a few nanoseconds width. The thermal load of the peak current to the driver MOS limits the output duty cycle to \SI{10}{\percent} on timescales longer than a few seconds. To protect the amplifier from spurious back-scattering its output is first passed through a two-stage Faraday isolator (Qioptiq FI-780-5TVC, \SI{75}{\deci\bel} isolation, \SI{85}{\percent} transmission) and as an abundance of caution two interference filters (Laseroptic L-12490, \SI{0.4}{\nano\meter} FWHM pass-bandwidth at \SI{795}{\nano\meter}) are used to filter out any broad amplified spontaneous emission. This setup is sketched in Fig.~2(a). 
	
	\begin{figure}
		\includegraphics[width=\columnwidth]{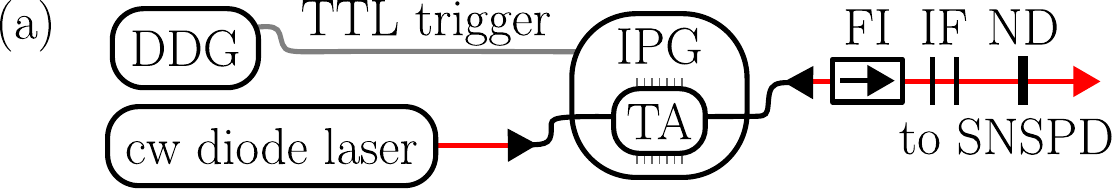}\\\vspace{0.2cm}
		\includegraphics[width=\columnwidth]{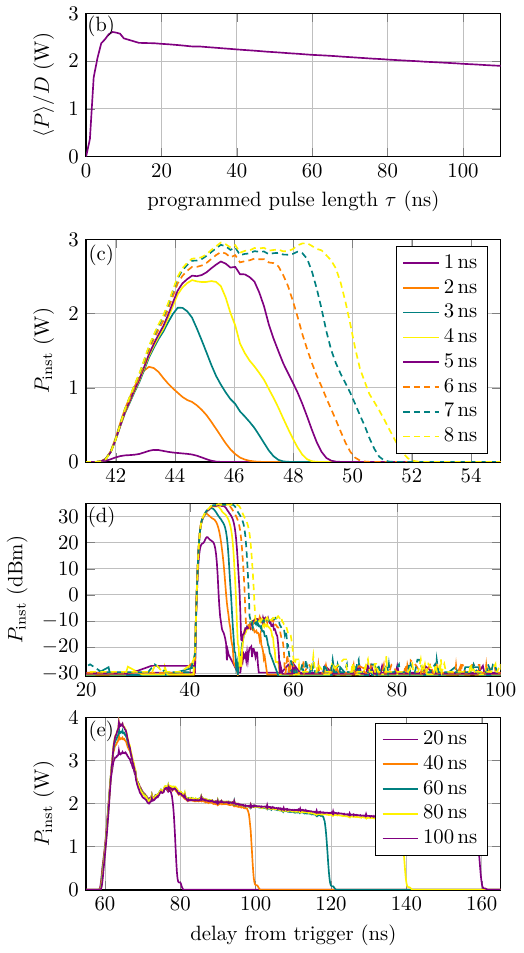}
		\caption{(a) Setup to characterize TA switching performance. DDG digital delay generator, TTL transistor-transistor logic, IPG current pulse generator, TA tapered amplifier, FI Faraday isolator, IF interference filters, ND neutral density, SNSPD superconducting nanowire single photon detector. (b) Time averaged optical power output of the TA under \SI{4}{\ampere} current pulsing at \SI{1}{\mega\hertz}, divided by the programmed pulse duty cycle, in \SI{1}{\nano\second} (\SI{2}{\nano\second}) steps for pulses up to (longer than) \SI{10}{\nano\second}. The rated cw output is \SI{2}{\watt}. All power measurements are accurate to \SI{5}{\percent}. (c) Instantaneous output power as a function of time and pulse length (see legend) in $2\times\SI{81}{\pico\second}$ time bins. (d) as (c) but on a logarithmic scale. Slight ringing after the pulses, four orders of magnitude below the peak amplitudes, is visible on timescales similar to the pulse length, whereupon the steady state IER is reached. (e) Behavior for longer pulse duration, revealing short term overshooting on mildly sloped rectangular pulse shapes.}
	\end{figure}
	
	The instantaneous power available within a pulse is generally the primary factor in determining achievable Rabi frequencies with which to drive an optical transition. As suitable lenses are simple objects, appropriate intensities are almost always obtainable and the power therefore sets the maximum mode volume addressable by optical control. This figure of merit is obtained in two steps. Short pulses of programmed length $\tau$ are generated every $T=\SI{1}{\micro\second}$. A slow-integrating thermal power sensor (Thorlabs S401C, \SI{5}{\percent} measurement uncertainty) is used to determine the average output power. The time averaged powers follow the naively expected relation of \SI{2}{\watt} specified continuous output at \SI{4}{\ampere} current times the programmed pulse duty cycle $D=\tau/T$ quite closely, data are shown in Fig.~2(b). Notable is the overshoot on specification for short pulses.
	
	Time resolved photon rate data from the SNSPD yield the pulse shapes with \SI{162}{\pico\second} time resolution and relative amplitude information over six orders of magnitude. To avoid errors stemming from detector dead time, average detection rates are lowered to $<\SI{1e5}{\per\second}$. This ensures accurate amplitude resolution on short term ringing at the cost of dynamic range, a worthy trade-off as the steady state IER of the TA is only \SI{58}{\deci\bel} and thus relatively easily resolvable. Short time switching behavior is shown in Figs.~2(c) and 2(d). These data are normalized to agree with the average power, yielding the absolute amplitudes after isolation and filtration to \SI{5}{\percent} accuracy.
	
	 A comparison of the few nanosecond scale pulses to longer ones, \SIrange{20}{100}{\nano\second} range shown in Fig.~2(e), reveals fortuitous abnormal behavior of the amplifier on short time scales. As inversion builds, the amplifier output rises to a maximum value in about \SI{3}{\nano\second} ($10:90$), but as foreshadowed by the average powers and confirmed by longer pulses, this initial peak is a considerable overshoot of its specified output and is maximally maintained for about \SI{5}{\nano\second}. The driver is designed to limit the peak current supplied to the TA and indeed should not be capable of passing more than the \SI{4}{\ampere} rated maximum through the TA by design. Limited control over the driver's behavior is possible via the compliance voltage, but adjusting this value has had no effect on extra optical output. As the output is into free space optical damage is of no additional concern, but naturally some general caution, e.g. with fiber end facets, is advisable at these powers. Note that the overshoot can be eliminated by a more conservative choice of set current at the price of lower peak and average output powers, however, as high peak powers are desirable, I have chosen to operate this TA as is over multiple months without finding performance degradation. On the long timescales shown in Fig.~2(e) it becomes clear that the initial overshoot is part of a damped ringing. This is, in fact, the expected response of a diode to a current step. For small signals the expected timescale is that of the photon lifetime (picoseconds), but here the signals are not small and gain compression makes the much longer charge carrier lifetime (nanoseconds) determine the temporal behavior \cite{Petermann1988}. One notable side effect is that the falling flank of the optical pulses follows two time scales, one set by the return from overshoot to the expected output power, and one corresponding to the current switching off. On short pulses these overlap and produce a roughly piecewise-linear falling flank. On long pulses, however, they are temporally well separated and can thus be cleanly specified to about \SI{3}{\nano\second} ($90:10$) and \SI{1.5}{\nano\second} ($90:10$) respectively. The latency of the device, defined and measured as above, is \SI{41}{\nano\second} for short pulses, increasing by a further \SI{17}{\nano\second} for $\tau\geq\SI{10}{\nano\second}$. This device has been integrated to upgrade multiple ongoing memory experiments, and has eliminated the significant limitation found in \cite{Buser2022,Mottola2023} of unintentional spin wave readout during the storage time.
	
	\section{RF Pulsing Laser Diodes for Chirping}
	
	We now turn to the question whether this current modulation approach can be profitably applied to establish laser phase control as well. For a single mode diode laser to emit light at a specific, stable frequency, the current passing though it should be constant. Therefore, to ensure reliable operation for all applications requiring spectroscopic accuracy, cw diode lasers are typically driven by current sources. These drivers rapidly vary the diode voltage, often within certain limits for over voltage protection, to maintain a stable set current running through it. On long time scales, such as those at play when adjusting the set current by hand via a potentiometer, frequency shifts of \SI{100}{\mega\hertz} can correspond to current changes as small as \SI{1}{\milli\ampere}. Devices capable of adding a nanosecond scale current pulse of such an amplitude onto a constant current two orders of magnitude larger fall well outside of typical design specifications. Pulsed laser drivers usually strive to reach peak currents within \SIrange{0.1}{1}{\ampere}, resulting in poor resolution at such low amplitudes. Moreover, with some exceptions once again born from telecom applications, they typically cannot drive the \SI{50}{\ohm} loads of standard bias-tees, complicating the addition of the pulse onto a constant current.
	
	In comparison to the regulatory timescale of a constant current source, a frequency chirp of a few nanoseconds effectively requires a fast transient effect on the instantaneous diode current. Intuitively, it is equivalent to achieve this by modifying the voltage across the diode much faster than the driver responds to stabilize the current. I pursue this approach by connectorizing a \SI{9}{\milli\meter} TO-can laser diode (Toptica LD-0790-0120-AR-3) in a home built interference filter based external cavity laser housing to a bias-tee (Minicircuits ZFBT-4R2GW-FT+). I run the output of a standard analog laser current driver (Toptica DCC 110) through the bias-tee's DC port while applying fast voltage signals from a pulse generator (Picoquant PPG512, \SI{200}{\pico\second} horizontal, \SI{8}{\bit} vertical resolution, \SI{12}{\volt} maximum amplitude into \SI{50}{\ohm}) to the rf port. As diodes respond very differently to fast transients than in cw\cite{Petermann1988}, the resultant effect is ill-suited to quantitative \emph{a priori} predictions. I proceed with an optical characterization of this ``chirp laser''.
	
	\begin{figure}
		\includegraphics[width=\columnwidth]{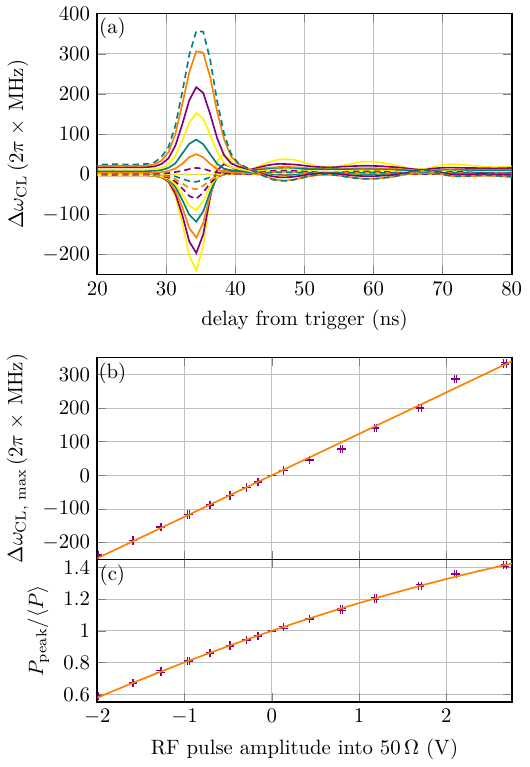}
		\caption{(a) Frequency chirps closely follow the \SI{5}{\nano\second} FWHM Gaussian shapes programmed onto the applied rf pulses. The amplitudes must be calibrated by this kind of optical characterization initially, but thereupon quite arbitrary chirps can be generated with minimal effort. (b) The chirp amplitude scales linearly with the amplitude of the applied rf pulse. The orange line is a fit to the slope alone yielding a scaling of \SI{123(2)}{\mega\hertz\per\volt}. (c) The laser intensity is changed in correlated and rather dramatic fashion by the rf pulses. The y-axis represents the ratio of laser power at the peak of the chirp to its average. At the standard operating current of \SI{230}{\milli\ampere}, chosen to allow demonstration of the technique's bidirectionality, $\langle P\rangle$ is \SI{69}{\milli\watt} at the laser head output and \SI{49}{\milli\watt} before the fiber to the amplifier. As shown by the fit line, the trend with rf amplitude is very mildly quadratic.}
	\end{figure}
	
	The chirp laser is offset locked at a frequency of $\omega_c=2\pi\times\SI{1.261}{\giga\hertz}$ to an auxiliary laser, which is itself locked to a rubidium line. The lasers are interfered on a $50:50$ beam splitter and one output is detected by a biased photo diode (Hamamatsu PD G4176-03, \SI{30}{\pico\second} rise and fall times). The resulting beat note is registered by a frequency counter (aim-TTi TF930) and is used to digitally generate feedback to a piezo modifying the laser cavity length using a computer run PI-control LabVIEW routine and integrated analog I/O card (NI PCIe-6363). Its long term frequency is thus stabilized to $\omega_{\text{CL,set}}=2\pi\times\SI{377.110526(1)}{\tera\hertz}$. For characterization purposes this is an arbitrary choice, important is only that $\omega_c$ is sufficiently larger than the frequency chirps to be measured. Simultaneously, the beat note between the lasers is monitored from the second beam splitter output with a fiber coupled photo detector (Thorlabs DXM25CF, \SI{25}{\giga\hertz} bandwidth) and oscilloscope (Tektronix DSO6, \SI{2.5}{\giga\hertz} bandwidth), and the diode's intensity response is recorded separately using a more economical detector (Thorlabs DET025AFC/M). The setup for this measurement is included in Fig.~4. 
	
	The challenge at hand is simultaneously achieving high time and frequency resolution near the Fourier limit. Even without applying rf pulses, the standard deviation on the beat signal, stabilized to $\omega_c$ but recorded in the time domain in a \SI{200}{\nano\second} trace, is about \SI{15}{\mega\hertz}. This is true even when a pure tone from a signal generator is measured in this manner and is merely a limitation of sampling, without corresponding to any real fluctuation. To reveal induced shifts on nanosecond timescales I proceed as advocated for by White et al.\cite{White2004}, recording such short time traces of the beat note, then fitting sine waves to the data in \SI{1}{\nano\second} windows. As the phase of the beat note with respect to the timing of chirping rf pulses is random, averaging in hardware merely erases the signal. Thus, to overcome statistical error, these frequency fits are averaged in post-processing over 12500 traces enabling sub-megahertz precision.
	
	Both the amplitude and frequency response of the diode functionally follows the applied rf pulses near perfectly. As a test case, I use Gaussian voltage pulses of \SI{5}{\nano\second} FWHM and vary their amplitude. Pulse width and flank form are imprinted onto the diode output. I forego an attempt to directly measure instantaneous electrical signals across the diode, orienting us by its optical response instead. I thus limit the amplitude of the rf pulses such that the maximum rating of the diode's optical output is not exceeded at their peaks, using weak attenuation in the rf line to make full use of the pulse generator's vertical resolution. The resulting chirps are shown in Fig.~3(a), with $\Delta\omega_{\text{CL}}=\omega_{\text{CL}}-\omega_{\text{CL,set}}$. Each trace in Fig.~3(a) corresponds to a single point in Fig.~3(b), where they are identified by their amplitudes. The ringing visible after higher amplitude pulses corresponds exactly to what is present in the electronic signal when directly measured, i.e. it is not a result of how the rf pulses are applied to the diode. A trace without a programmed pulse is included as well -- here the frequency determined by fitting and averaging varies (peak-to-peak) less than \SI{500}{\kilo\hertz} from $\omega_{\text{CL,set}}$ over its duration. The corresponding intensity responses of the diode are shown in Fig.~3(c). As these are comparatively extreme and simultaneously easily measured they serve as an excellent proxy for fine adjustments to the chirp amplitude. Moreover, to gain insight into the single shot repeatability (systematic error) of the chirp amplitude precluded by this frequency measurement technique, I measure the shot-to-shot variation in intensity response. Its standard error, from pulsing over a few minutes and 40000 shots, converges to \SI{0.45}{\percent} variation at 1$\sigma$. For the largest demonstrated chirp amplitudes this equates to a single shot frequency accuracy slightly better than \SI{2}{\mega\hertz}. 
	
	The rf pulse amplitudes as measured into \SI{50}{\ohm} that produce these chirps are the x-axis of Fig.~3(b) and (c). The data reveal an effectively Ohmic frequency response of the diode to the applied transient. This linear behavior illuminates what occurs in the diode electrically -- although diodes are non-linear devices the applied pulses leave this one firmly in the linear regime. The effective change in current across it must therefore be small, akin to the $\leq\SI{1}{\milli\ampere}$ changes required to induce such laser frequency shifts in cw operation. This also explains why these pulses do not induce mode hops in the laser and can be run stably and continually all day long. At a glance, this is at odds with the intensity response seen in Fig~3(c), which is comparatively dramatic; peak powers readily increase by 10s of milliwatt, and care must even be taken not to exceed absolute maximum ratings of optical power. Were the modulation slow, such changes would require correspondingly significant changes to the current, but not so on the short time scale at play here as strong output overshoot is expected in the current step response limit \cite{Petermann1988}.
	
	The positive and negative amplitude pulses are generated by different generations of the same nominative model of pulse generator. The slight difference in the temporal shape of positive and negative chirps can be traced back to differences in the electronic pulse shapes of these two devices. This is a choice of convenience as while both are unidirectional in amplitude one happens to have been designed with inverted output. One such device plus a sufficiently fast voltage inverter should yield an identical effect. In addition to some ringing after the chirp, an amplitude dependent shift to the baseline laser frequency is induced by positive voltage pulses despite frequency stabilization. The scale of this shift is roughly \SI{8}{\percent} of the amplitude. As only one pulse generator produces this effect, it is most probably an electronics problem. Fortunately, it can be compensated for trivially by modifying the offset lock frequency. For the steepest flanks producible by these pulse generators (programmed rectangle) the effect on the optical output ($10:90$) is achieved in \SI{1.8}{\nano\second}, corresponding to a maximum chirp rate of about \SI{150}{\mega\hertz\per\nano\second} at the greatest tested rf amplitude. For this and any similar system this limit is ultimately a function of risk tolerance, as eventually the rf power enters a regime where diode damage becomes likelier. It is worth noting that, although analyzing chirps by measuring changes in a beat note is laborious, once the proportionality of the induced intensity and frequency shifts on the timescale of interest is established, it is theoretically sound to extrapolate the effect of similar duration voltage pulses on laser phase by simply measuring amplitude variations. This drastically mitigates the effort required for adjustments after an initial setup.
	
	\section{Chirped and Amplified Laser Pulses}
	
	Combination of the techniques presented so far to produce pulses with instantaneous power amplitudes of \SI{3}{\watt} and arbitrarily shaped chirps within a few hundred megahertz is as simple as using the chirp laser to optically seed the TA, then electronically aligning the rf pulse for chirping to the current pulse for amplifying in time, accounting for insertion and propagation delays. A complete setup of the combined system is sketched in Fig.~4. As both amplitude and frequency vary dramatically over the short duration of an amplified and chirped pulse, and as there is no measurable reference beat note outside of the amplification time, frequency analysis as performed above suffers from technical artifacts and is accurate when performed with automated routines befitting the data volume only over the very short durations for which the pulses are fully on. Therein the data agree with the characterization shown in Fig.~3(a), and short chirp pulses on top of longer stretches of amplification also reproduce the data of the previous section. Including all device and propagation induced delays, the output can be produced in less than \SI{65}{\nano\second} upon receiving an asynchronous external trigger. 
	
	The combined system has been integrated into the latest iteration of the memory experiment initially described in \cite{Buser2022}, and some improvements to the memory efficiency have already been achieved. That this is possible and advantageous is a result of two major distinguishing features over state-of-the-art external modulation approaches to generating chirped and amplified light such as that by Clarke and Gould \cite{Clarke2022}, chosen for comparison here both due to its overall impressive performance and as it represents typical attributes of external modulation systems. The first is the asynchronous responsiveness of the output, as the external system relies on a series of 4 AOMs it is principally incompatible with experiments that require a full response to an asynchronous trigger within 10s of nanoseconds -- the external approach is at least an order of magnitude too slow. The second is the intensity extinction shortly after a pulse. This attribute is non-negotiable to a single photon experiment, but an entirely ancillary concern in \cite{Clarke2022} to the point where it is hardly specified. From amplitude data shown on a linear scale in their Fig~3, we can estimate that it is worse than \SI{-20}{\deci\bel}, and from the typical performance of devices used it is safe to say that \SI{-30}{\deci\bel} cannot be exceeded on timescales comparable to the pulses themselves using their or a similar approach.
	
	\begin{figure}
		\includegraphics[width=\columnwidth]{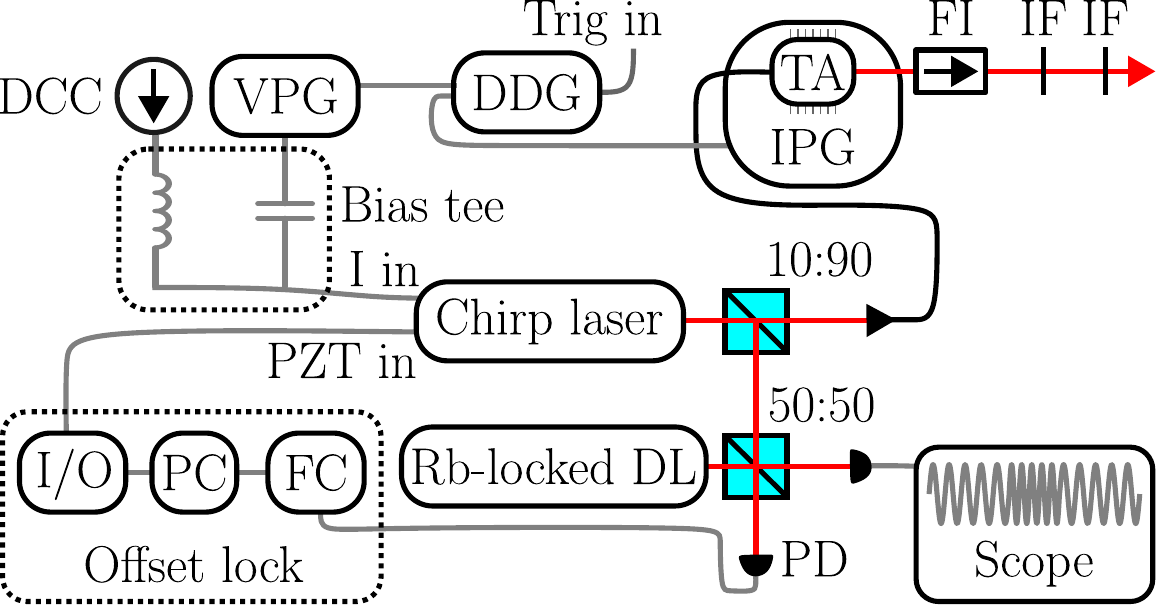}
		\caption{Experimental setup to produce and characterize chirped and amplified laser pulses, on-demand and entirely with direct current modulation. DDG digital delay generator, VPG voltage pulse generator, DCC constant current source, I in laser current input, X:Y beam splitter with ratio R:T, DL diode laser, PD photo detector, FC frequency counter, PC computer running PI control software, I/O analog input/output card, PZT in laser piezo input, IPG current pulse generator, TA tapered amplifier, FI Faraday isolator, IF interference filter.}
	\end{figure}
	
	Amplification of chirped pulses has a significant advantage beyond just reaching higher peak powers. The TA saturates from \SI{20}{\milli\watt} input, and it tolerates up to \SI{50}{\milli\watt} of injected seed light. Thus, there exists an operating regime where it is saturated for all instantaneous input powers present during a chirp. This tempers undesired intensity modulation on the output pulse. The maximum input power variation resulting from chirping produces less than \SI{2}{\percent} output power variation from the TA, tending towards zero with increasing pulse length. Implemented with an amplifier suited for cw operation, this approach would decouple phase and amplitude control despite relying on direct current modulation. Moreover, in the typical use case of optical pulses where desirable chirp coincides with intensity variation, ringing in the chirp signal is made irrelevant as the amplitude modulation effectively gates it away.
	
	\section{Applications, Limitations, and Outlook}
	
	High power laser pulses are an important tool of quantum control. Generally speaking, they are used to drive transitions, e.g. Raman transitions in atom optics. Optimal control often simultaneously demands that these pulses exhibit some chirp, for instance to maintain adiabaticity conditions \cite{Collins2012} or tailor forces on atoms \cite{Metcalf2017}. As a particularly illustrative example as well as the initial motivation for this investigation, consider the application of a quantum memory for spectrally broad single photons\cite{Fleischhauer2002,Gorshkov2007}. An ensemble of three-level lambda systems has 2 metastable ground states which we label $\ket{s}$ and $\ket{g}$. The systems are prepared in $\ket{g}$. An incoming signal is near resonant to the transition from the prepared ground to the third, excited state $\ket{g}$--$\ket{e}$. Ideally, a strong control pulse on the $\ket{e}$--$\ket{s}$ transition and in two-photon resonance with the signal can now be used to map the photon into a spinwave excitation between the two ground states. In the appropriate limit for fast processes, the bandwidth of this memory scheme is roughly equal to the Rabi frequency of the control, which in turn is linear in its amplitude. The problem with a fixed frequency control pulse is apparent -- as its peak amplitude is necessarily high for a broadband signal, the pulsed control induces a time varying light shift on $\ket{s}$, precluding the optimal storage conditions of continuous two-photon resonance. This effect is known to limit memory efficiencies in high bandwidth, off-resonant storage schemes \cite{Buser2022,Davidson2023,Mottola2023}. Critically, to interface such memories with probabilistic photon sources or to make use of them in realistic networking applications requires that all of their components rapidly respond to an asynchronous external trigger. This is typical of technological applications.
	
	For this class of experiments the current-based approach to chirping and amplifying shines particularly brightly. It is conceptually and technologically simple and exceptionally performant for its price point. It offers less flexibility in amplitude shaping and cannot reach the same chirping range as state-of-the-art external modulation systems\cite{Clarke2022}, not even by ignoring absolute maximum ratings. To demonstrate the ability of this technique to chirp in either direction, I have operated the laser diode at an intermediate set current, giving plenty of room in both directions before optical damage becomes a concern. If only positive (negative) chirps are required, the diode could instead be operated near its lasing threshold (maximum current) to yield approximately double the maximum amplitude and speed, but this does not fundamentally change the preceding conclusion. Moreover, it demands some straightforward optical characterization of the performance for every diode the technique is implemented with, which scales only tediously. Nevertheless, all characterizations here have yielded neat and predictive phenomenology for reasonable effort. In exchange, switching, pulsing and chirping are all achieved rapidly upon demand and with vastly superior intensity extinction than any external modulation system matching the speed and power handling. These attributes make this approach the prime candidate for experiments that have been treading the most common path of quantum technological investigations where optimal control requires chirping, which is just to forgo optimal control. Similarly, many experiments with photons limited by control-induced noise or hampered in their rates by required filters can benefit from the high intensity extinction ratios demonstrated here, likely even by using already commercialized devices.
		
		\begin{acknowledgments}
			I cordially thank Philipp Treutlein for feedback on the manuscript and probing questions pushing the quality of my device characterizations, Michael Steinacher for his design of the SOA driver and helpful conversations regarding the rf pulsing of the laser diode, Roberto Mottola for obtaining the butterfly packaged TA and high current driver, including a custom solution to an atypical diode pin layout, and Janik Wolters, who has always claimed that atomic physics should borrow ideas from the telecom realm with more enthusiasm. Further, I gratefully acknowledge financial support from the Swiss National Science Foundation and the Swiss State Secretariat for Education, Research, and Innovation (SERI) through the project Scalable High Bandwidth Quantum Network (sQnet).
		\end{acknowledgments}
		
		\section*{Data Availability Statement}
		
		The data that support the findings of this study are available from the corresponding author upon reasonable request.
		
		\bibliography{switches}
		
	\end{document}